\documentclass[%
 reprint,
 amsmath,amssymb,
 aps,
]{revtex4-1}

\usepackage{amssymb}   
\usepackage{amsmath}   

\usepackage{graphicx}
\usepackage{hyperref}
\usepackage{float}
\usepackage{color}

\usepackage{graphicx}
\usepackage{dcolumn}
\usepackage{bm}

\begin{document}

\preprint{UFES}

\title{Degeneracy between nonadiabatic dark energy models and $\Lambda$CDM: ISW effect and the cross correlation of CMB with galaxy clustering data}

\author{Hermano Velten$^{1}$}
 \altaffiliation[]{velten@pq.cnpq.br (corresponding author)}
\author{Raquel Emy Fazolo$^{1}$}%
\author{Rodrigo von Marttens$^{1,2}$}%
\author{Syrios Gomes$^{1}$}%
\affiliation{%
$^{1}$N\'ucleo COSMO-UFES \& Departamento de F\'isica,  Universidade Federal do Esp\'irito Santo (UFES)\\
 Av. Fernando Ferrari s/n CEP 29.075-910, Vit\'oria, ES, Brazil }%
\affiliation{%
$^{2}$Departamento de Astronomia, Observat\'orio Nacional, 20921-400, Rio de Janeiro, RJ, Brasil}%
\date{\today}

\begin{abstract}
As recently pointed out in Ref. [Phys.\ Rev.\ D {\bf 96}, 8, 083502 (2017)] the evolution of the linear matter perturbations in nonadiabatic dynamical dark energy models is almost indistinguishable (quasi-degenerated) to the standard $\Lambda$CDM scenario. 
In this work we extend this analysis to CMB observables in particular the integrated Sachs-Wolfe effect and its cross-correlation with large scale structure. We find that this feature persists for such CMB related observable reinforcing that new probes and analysis are necessary to reveal the nonadiabatic features in the dark energy sector.
\begin{description}
\item[PACS numbers]
\end{description}
\end{abstract}

\maketitle


\section{\label{sec:level1}Introduction}

The modern view of the cosmological evolution is based on the use of general relativity (GR) as the standard gravitational theory equipped with baryonic matter, neutrinos and photons as well as additional nonstandard components named dark matter and dark energy. Concerning the role played by such constituents of the universe energy budget on the cosmological evolution there is another standard assumption related to their thermodynamical nature. They are assumed to be adiabatic fluids with well established equations of state. Cold dark matter and baryons are assumed to be pressureless ($p_B=p_{DM}=0$), due to the relativistic nature of photons $p_{\gamma}=\rho_{\gamma}/3$ (where $\rho_{\gamma}$ is the energy density of photons), neutrinos suffer a transition from the relativistic regime to the non-relativistic one and dark energy is characterized by a negative equation of state parameter $w_{DE}=p_{DE}/\rho_{DE}$. Being the latter fact necessary to account for the accelerated background expansion rate experienced on late times. 

It is worth noting that this overall picture is also consistent when the structure formation process is analysed. However, in cosmological perturbation theory there exists another degree of freedom related to the possible nonadiabatic nature of the cosmic components. The possible nonadiabatic features manifest either as a intrinsic property of each fluid (intrinsic entropy perturbations) or due to the multi-fluid nature of the system (also known as relative entropy perturbations). Standard cosmological scenario does not take into account any of such effects. 

It has been found recently in Ref. \cite{Velten:2017mtr} that nonadiabatic dark energy models are almost indistinguishable from the standard $\Lambda$CDM cosmology concerning the $f \sigma_8$ observable leading to an interesting degeneracy in the dark energy research context.
Our focus in this work is to expand the analysis of Ref. \cite{Velten:2017mtr} by studying whether or not such feature persists if other probes are considered. In particular we study the evolution of the gravitational potential which is related to the so called integrated Sachs-Wolfe effect (ISW) and its cross correlation with matter clustering.

The dark energy models adopted in this work are presented in the next sections \ref{sec:background} and \ref{sec:de}. Each model has its specific equation of state parameter leading to the corresponding adiabatic speed of sound. In section \ref{sec:perturbations} we establish the first order equations of motion taking into account scalar perturbations only. Working in the Newtonian gauge the perturbations are characterized by the potentials $\Psi$ and $\Phi$, but in case of vanishing anisotropic stresses sourced by the energy-momentum tensor they coincide i,e., $\Psi=\Phi$. We adopt it in this work. The time evolution of the Newtonian potentials is related to a secondary source of temperature anisotropies in the Cosmic Microwave Background (CMB) called integrated Sachs-Wolfe (ISW) effect. Here, since we are focus on the dynamics of dark energy models we investigate the late time ISW effect \footnote{The early ISW effect occurs around the decoupling time and is caused by a effective residual radiation/neutrino contribution to the cosmological background dynamics.}. During the matter dominated epoch the clustering of matter within the potential wells is exactly counterbalanced by the background expansion leading to a constant $\Phi$ behavior. The ISW effect acts on large scales (low-$l$ multipoles in the CMB power spectrum) and is difficult to be independently detected due to its tiny contribution (around one order of magnitude smaller) to the averaged temperature fluctuations. Since it is caused by the evolution of potential wells in an expanding universe one possible strategy is to cross-correlate the CMB map with galaxy distribution maps from large scale structure (LSS) data. This procedure gives rise to the so called temperature-galaxy cross-spectrum (the $C^{Tg}_l$ spectrum). See for example Ref. \cite{Ho:2008bz}.

We study in detail the ISW effect for adiabatic and nonadiabatic dark energy cosmologies. One of the perspectives we adopt here follows the analysis employed by Ref. \cite{Dent:2008ek} in which the relative amplification of the ISW effect of a dark energy model is computed in comparison to the standard $\Lambda$CDM model. This analysis is performed in Section \ref{Qsection}. The cross-correlation spectrum for nonadiabatic dark energy models and the comparison with the adiabatic cases is studied in section \ref{sec:Results}. We conclude in the final section \ref{sec:Conclusion}.

\section{\label{sec:level1}Dynamics of dark energy models}
\subsection{\label{sec:background} Background expansion}

Concerning the background evolution we are interested in the late time effects which are mostly caused by dark energy. At this stage the effects due to the radiation fluid are negligible. Indeed, the late time ISW effect is intrinsic to the dark energy phenomena. 

By adopting a flat Friedmann-Lemaitre-Robertson-Walker (FLRW) metric the expansion rate reads 
\begin{equation}
\frac{H^{2}(a)}{H^{2}_0}=\frac{\Omega_{\rm m0}}{a^{3}}+(1-\Omega_{\rm m0})\,\, e^{-3\int da\frac{1+w_{\rm DE}}{a}},
\label{Ha}
\end{equation}
where $H=\dot{a}/a$, $a$ is the scale factor and the symbol ``\,$\,^{.}\,$\,'' means derivative with respect to the cosmic time. For the reference $\Lambda$CDM model adopted in this work we fix the parameters 
$\Omega_{\rm m0}= 8\pi G \rho_{m0}/3H^2_0=0.2973$, $H_0=70.23 km \,s^{-1} Mpc^{-1}$ and $w_{DE}=-1$ (see the results for the statistical analysis presented in Table I). For other dynamical dark energy models the expansion rate (\ref{Ha}) is fully determined by the choice of the dark energy equation of state parameter $w_{DE}=p_{DE}/\rho_{DE}$.

\subsection{\label{sec:de}Dark energy models at background level}

There are some relevant parameterizations for $w_{DE}$. In this work (as done in Ref. \cite{Velten:2017mtr}), we will explore the following cases:
\begin{itemize}
\item The constant EoS parameter ($w$CDM) 
\begin{equation}
w_{\rm DE}=w_0 ;
\end{equation}
\item The Chevallier-Polarski-Linder model (CPL) \cite{Chevallier:2000qy,Linder:2002et}
\begin{equation}
w_{\rm DE}(a)=w_{0}+w_{1}(1-a);
\end{equation}
\item The Wetterich-logarithmic model (WL) \cite{Wetterich:2004pv}
\begin{equation}
w_{\rm DE}(a)=\frac{w_{0}}{\left[1+w_{1}{\rm ln}(1/a)\right]^2}.
\end{equation}
\end{itemize}

\subsection{\label{sec:perturbations} Scalar perturbations of dark energy models}

Following refs. \cite{Malik:2002jb, Bartolo:2003ad, Copeland:2006wr, Dent:2008ek}, the line element including scalar perturbations up to first order is written according to
\begin{align}
ds^{2}=-(1+2A)dt^{2}+2a\partial_{i}Bdx^{i}dt \nonumber \\ \qquad
+a^{2}[(1+2\psi)\delta_{ij}+2\partial_{ij}E]dx^{i}dx^{j},
\end{align}
where the scalar metric perturbations are represented by the functions $A, B, \psi$ and $E$. By adopting the Newtonian (or longitudinal) gauge, i.e., $E=B=0$, the metric perturbations are written in terms of the so called Newtonian potentials $\Phi$ and $\Psi$:
\begin{equation}
A-\frac{d}{dt}[a^{2}(\dot{E}+B/a)]\rightarrow A \equiv \Phi,
\end{equation}
\begin{equation}
-\psi+a^{2}H(\dot{E}+B/a)\rightarrow -\psi \equiv \Psi.
\end{equation}

The strategy we adopt here is to assume the entire cosmic energy budget (Cold Dark matter + Dark energy) as a single effective total fluid with density $\rho$ and pressure $p$. With this choice, the components of the energy momentum tensor of such one-fluid description reads 
\begin{align}
\tensor{T}{^0_0}=-(\rho+\delta\rho), \quad \tensor{T}{^0_\alpha}-(\rho+p)v_{,a}, \nonumber \\
\tensor{T}{^\alpha_\beta}=(p+\delta p)\delta^{\alpha}_{\beta}+\Pi^{\alpha}_{\beta}.
\end{align}
Since we neglect \textbf{the} anisotropic stresses ($\Pi^{\alpha}_{\beta} =0$) both Newtonian scalar potentials coincide $\Phi=\Psi$. With this condition one can compute the components of the Einstein's equation. They read
\begin{equation}
-\frac{\nabla^2}{a^{2}}\Phi+3H^{2}\Phi+3H\dot{\Phi}=-4\pi G\delta\rho,
\end{equation}
\begin{equation}
H\Phi+\dot{\Phi}=4\pi Ga(\rho+p)v,
\end{equation}
\begin{equation}
3\ddot{\Phi}+9H\dot{\Phi}+(6\dot{H}+6H^{2}+\frac{\nabla^2}{a^{2}})\Phi=4\pi G(\delta\rho+3\delta p),
\end{equation}
where we have defined the fluid velocity potential as $\delta u^{i}\equiv \partial^{i}v$, where $u^{i}$ is the fluid four-velocity. Other necessary equations to determine the perturbative sector are obtained
from the covariant conservation of the energy momentum tensor
\begin{equation}
\delta\dot{\rho}+3H(\delta\rho+\delta p)=(\rho+p)(3\dot{\Phi}+\frac{\nabla^2}{a}v),
\end{equation}
as well as the momentum conservation
\begin{equation}
\frac{[a^{4}(\rho+p)v]^{\cdot}}{a^{4}(\rho+p)}=\frac{1}{a}\Big{(}\Phi+\frac{\delta p}{\rho+p}\Big{)}.
\end{equation}

In order to assess astrophysical scales of interest we work in the Fourier space  in which $\nabla^2 \rightarrow -k^2$. From the above equations the total density contrast $\Delta = \delta \rho / \rho$ can be written in terms of $\Phi$ such that
\begin{equation}
\Delta=-\Big{(}\frac{2k^{2}}{3a^{2}H^{2}}\Big{)}\Phi-2\Phi
-2\frac{\dot{\Phi}}{H}.
\label{DeltaPhi}\end{equation}

The total equation of state parameter of the cosmic medium is defined as $w(a)=\sum\Omega_{i}(a)w_{i}(a)$. In practice, since we are considering a pressureless matter component $p_M=0$ we obtain
\begin{align}
w(a)&=\Omega_{\rm DE}(a)w_{\rm DE}(a) \nonumber \\ 
&=\Omega_{\rm DE0}\Big{[}\frac{H_{0}^{2}}{H^{2}(a)}e^{-3\int da\frac{1+w_{DE}}{a}}\Big{]}w_{DE}(a).
\end{align}

Then, it is convenient to use the sound speed definition $c_{s}^{2}\equiv \delta p/\delta\rho$, and to combine the above equations in order to write down the following result for the definition $\Theta = (c^2_{s} - w)\Delta$ such that
\begin{equation}
\Theta=\frac{2}{3H^{2}}[\ddot{\Phi}+H(4+3w)\dot{\Phi}+w\frac{k^{2}}{a^{2}}\Phi].
\end{equation}
The evolution of $\Delta$ in terms of the velocity potential becomes 
\begin{equation}
\dot{\Delta}=-3H\Theta+3(1+w)\dot{\Phi}-(1+w)\frac{k^{2}}{a}v,
\label{cont}
\end{equation}
and the dynamical equation for the velocity potential $v$ reads 
\begin{align}
\dot{v}=-vH(1-3w)-\frac{\dot{w}}{1+w}v+\frac{1}{a}\Big{[}\Phi+\frac{w}{1+w}\Delta \nonumber \\ \qquad+\frac{\Theta}{1+w}\Big{]}.
\label{Euler}
\end{align}

According to \cite{Bartolo:2003ad} intrinsic entropic perturbations in the dark energy fluid can be introduced via the definition
\begin{equation}
\Gamma(a)\equiv\frac{3H(1+w_{\rm DE})c^{2}_{\rm a,DE}}{1-c^{2}_{\rm a,DE}}\Big{(}\frac{\delta\rho_{\rm DE}}{\dot{\rho}_{\rm DE}}-\frac{\delta p_{\rm DE}}{\dot{p}_{\rm DE}}\Big{)}.
\label{Gammadef}
\end{equation}
In multi-fluid systems, as the one we are dealing here, it is also possible to define the relative entropic perturbations. For the system consisting of pressureless matter and dynamical dark energy this quantity becomes
\begin{equation}
S(a)\equiv\frac{3H(1+w_{\rm DE})\Omega_{\rm m}}{1+w}\Big{(}\frac{\delta\rho_{DE}}{\dot{\rho}_{\rm DE}}-\frac{\delta\rho_{\rm m}}{\dot{\rho}_{\rm m}}\Big{)}.
\label{Sdef}
\end{equation}

By decomposing the total pressure perturbation as $\delta p=\delta p_{nad}+c^{2}_{a}\delta\rho$, the above relations allows us to write the intrinsic part of the nonadiabatic pressure perturbation as
\begin{equation}
\delta p_{\rm nad}=\Omega_{\rm DE}[(-c^{2}_{\rm a,DE})S+(1-c^{2}_{\rm a,DE})\Gamma]\rho.
\end{equation}

The intrinsic adiabatic speed of sound of the total fluid is therefore written as
\begin{equation}
c_{\rm a}^{2}=\frac{\dot{p}}{\dot{\rho}}=\frac{w}{1+w}\Big{[}(1+w_{\rm DE})-\frac{a}{3}\frac{w^{\prime}_{\rm DE}}{w_{\rm DE}}\Big{]},
\end{equation}
where the symbol prime ``$\,^{\prime}\,$'' means derivative with respect to the scale factor.

Also, for the separated dark energy components its intrinsic dark energy adiabatic speed of sound reads
\begin{equation}
c^2_{\rm a, DE} = \frac{\dot{p}_{\rm DE}}{\dot{\rho}_{\rm DE}}=w_{\rm DE}-\frac{w_{\rm DE}^{\prime} a }{3(1+w_{\rm DE})}.
\end{equation}

Combining the above equations we are able now to produce a set of equations for the gravitational potential, the intrinsic entropic perturbation $\Gamma$ and the relative entropic perturbation $S$. These equations will be used for assessing the impact of nonadiabatic dark energy perturbation on the ISW effect. The full perturbative dynamics is then achieved after solving the following coupled equations for $\Phi, S$ and $\Gamma$,

\begin{align}
a^{2}H^{2}\Phi^{\prime\prime}+\Big{(}5aH+a^{2}H^{\prime}+3aHc_{a}^{2}\Big{)}a^{2}H \Phi^{\prime} \nonumber \\
\qquad +c_{\rm a}^{2}k^{2}\Phi+\Big{(}3+2a\frac{H^{\prime}}{H}+3c_{\rm a}^{2}\Big{)}a^{2}H^{2}\Phi= \nonumber \\
\qquad \frac{3}{2}a^{2}H^{2}\Omega_{\rm DE}\Big{[}-c_{\rm a,DE}^{2}S+\Big{(}1-c_{\rm a,DE}^{2}\Big{)}\Gamma\Big{]},
\label{Pot1}
\end{align}

\begin{align}
aS^{\prime}=\Big{(}3w_{\rm DE}-\frac{3\Omega_{m}c_{\rm a,DE}^{2}}{1+w}\Big{)}S \nonumber \\
\qquad +\frac{3\Omega_{\rm m}(1+c_{\rm a,DE}^{2})\Gamma}{1+w}+\frac{k^{2}}{a^{2}H^{2}}\frac{S+\Gamma}{3} \nonumber \\
\qquad+\frac{k^{4}}{a^{4}H^{4}}\Big{(}\frac{2}{9}\frac{1+w_{\rm DE}}{1+w}\Big{)}\Phi,
\label{Pot2}
\end{align}
\begin{align}
a \Gamma^{\prime}=-\frac{3}{2}(1+w)S+3\Big{(}w_{\rm DE}-\frac{1+w}{2}\Big{)}\Gamma \nonumber \\ 
\qquad +\frac{k^{2}}{a^{2}H^{2}}\Big{(}-(1+w_{\rm DE})\mathcal{R}-\frac{S+\Gamma}{3}\Big{)} \nonumber \\
\qquad +\frac{k^{4}}{a^{4}H^{4}}\Big{(}-\frac{2}{9}\frac{(1+w_{\rm DE})}{1+w}\Phi\Big{)}.
\label{Pot3}
\end{align}
In Eq. (\ref{Pot3}) we have defined the gauge-invariant comoving curvature perturbation 
\begin{equation}
\mathcal{R}\equiv \Phi+\frac{2}{3(1+w)}\Big{[}\Phi+a\frac{d\Phi}{da}\Big{]}.
\end{equation}

In the absence of nonadiabatic perturbations $S = \Gamma = 0$ the right hand side of Eq. (\ref{Pot1}) vanishes and the standard equation for the adiabatic cosmology is recovered.

\subsection{\label{Qsection}Evolution of the gravitational potential}

Now we promote a comparison between the adiabatic (AD) and the non-adiabatic (NAD) models. We call adiabatic dark energy model the potential $\Phi$ obtained solving Eq. (\ref{Pot1}) with vanishing right hand side i.e., $\Gamma=S=0$. For the NAD model we solve the coupled set of Eqs. (\ref{Pot1}) - (\ref{Pot3}). In this case the potential $\Phi$ is sourced by the functions $S$ and $\Gamma$. It is worth noting that the $k^4$ scale dependence present in Eqs. (\ref{Pot1}) - (\ref{Pot3}) represents a new feature introduced by the nonadiabatic effects which should be relevant for sub-horizon modes.

The ISW effect contribution to the CMB temperature fluctuation can be calculated via the integration of the time variation of the gravitational potential along of the photon trajectory from the decoupling time ($a_{d}$) to the present time ($a_0$). For each wavenumber $k$, this contribution reads

\begin{equation}
\left(\frac{\Delta T}{T}\right)^{ISW}_k = 2 \int^{a_0}_{a_d} \frac{\partial \Phi_k}{\partial a} da= 2\left( \Phi_k(a_0)-\Phi_k(a_{d})\right)
\label{DeltaTISW}\end{equation}
In the above equation we have neglected the photon optical opacity.

When solving equations Eqs. (\ref{Pot1}) - (\ref{Pot3}) the same adiabatic initial conditions are used i.e., $\Gamma(a_d)=S(a_d)=0$. Even for the nonadiabatic models this should be the case. Indeed, we study the evolution of late time nonadiabatic features.

We define the fractional excess $Q$ of the ISW effect produced by the (adiabatic or nonadiabatic) dark energy models with respect to the standard $\Lambda$CDM model according to
\begin{equation}
Q_{MODEL}= \frac{\left( \Delta T/ T \right)^{ISW}_{MODEL} } {\left( \Delta T/ T \right)^{ISW}_{\Lambda CDM}} - 1 
\end{equation}
Although the existence of cosmic variance in the CMB data, a viable cosmological model can not yield to a large amplification of the ISW effects. For example, this aspect has ruled out unified bulk viscous cosmologies \cite{barrow,Velten:2011bg}. 

For our first analysis we are going to set a specific scale of interest $k=0.0005 h Mpc^{-1}$ which is linear and large enough to suffer the main ISW contribution. 

\begin{figure}
\includegraphics[width=0.45\textwidth]{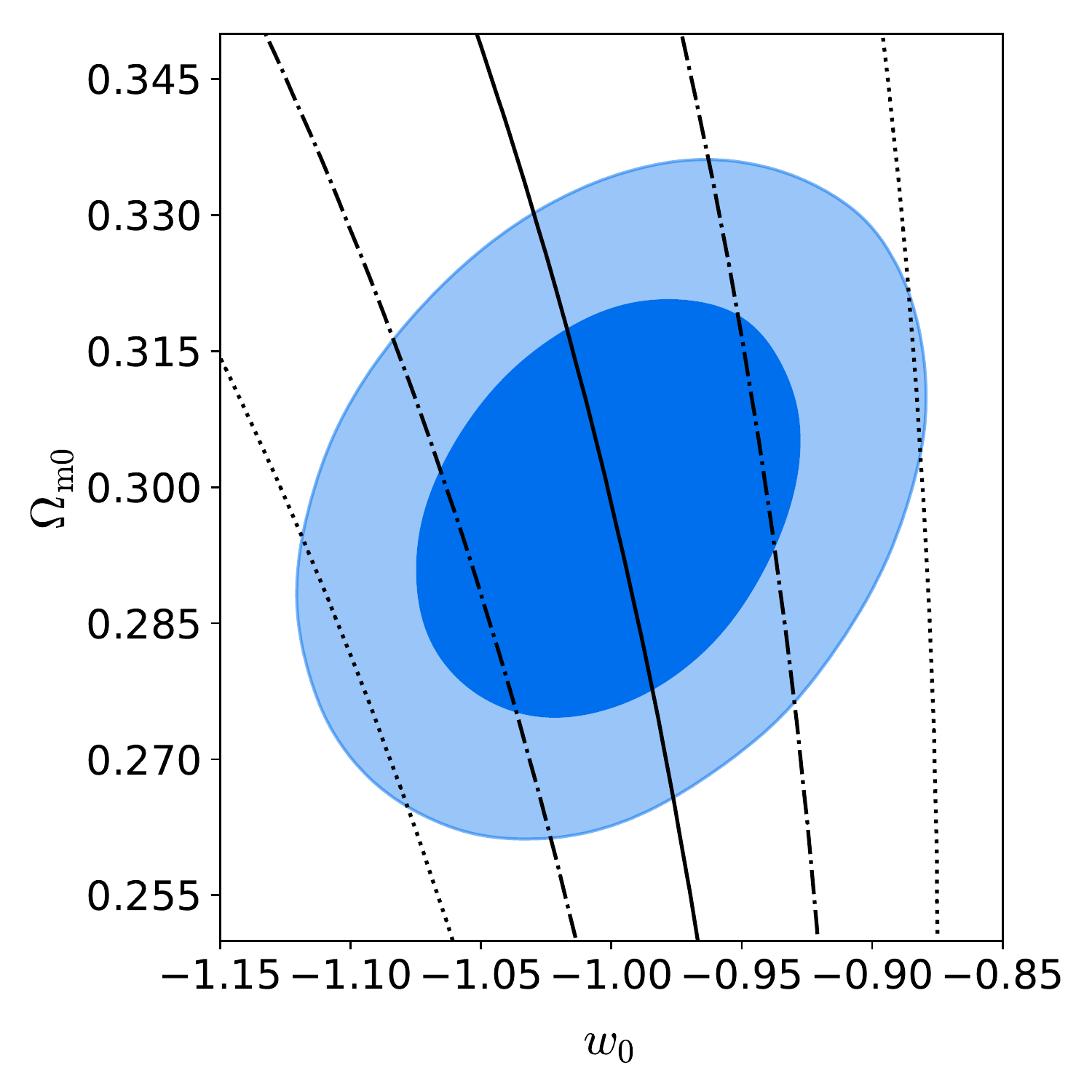}
\includegraphics[width=0.45\textwidth]{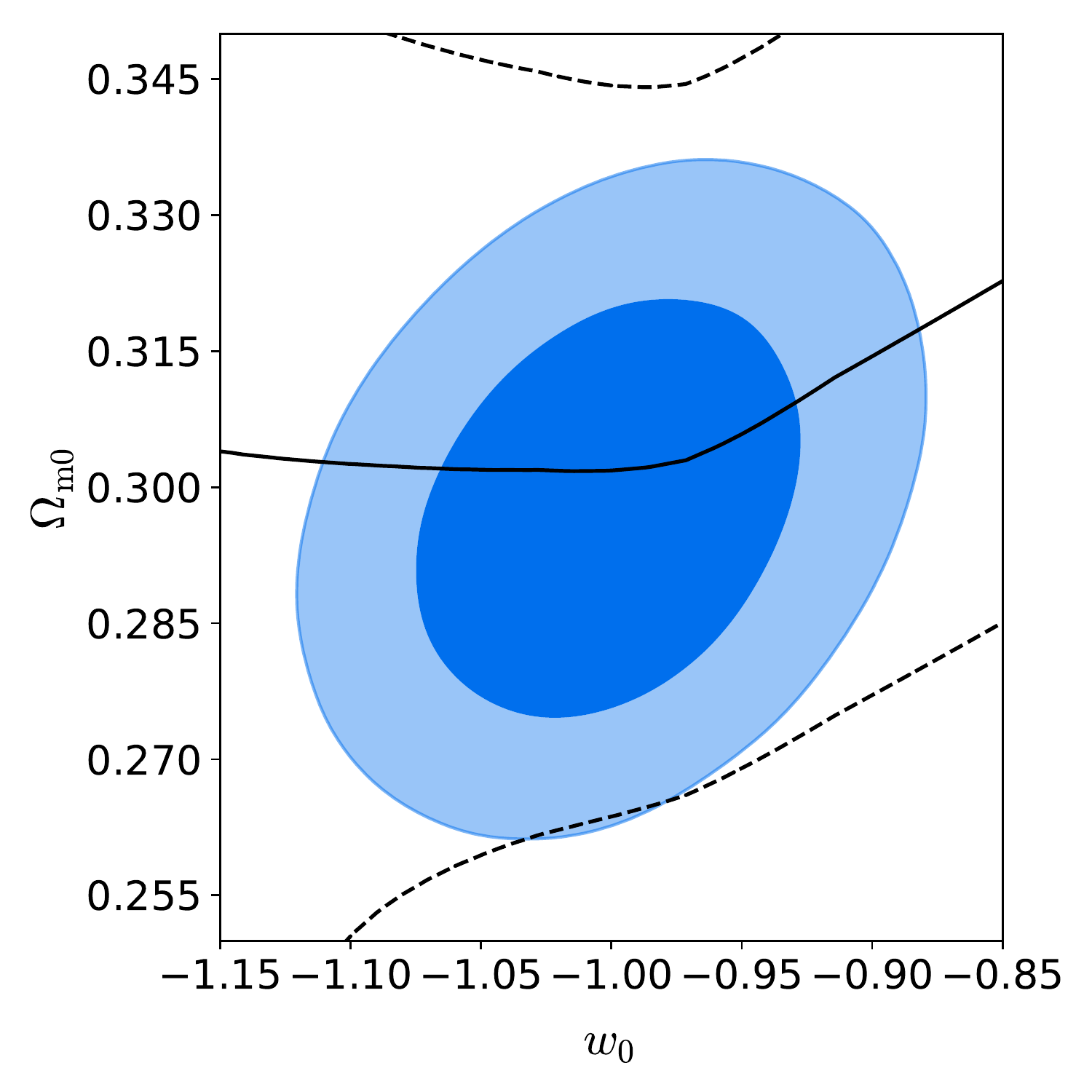}
\caption{Joint constraints (blue contours) and constant Q-values (black lines) in the plane $w_0$ x $\Omega_{m0}$ for the $w$CDM model. Top panel: adiabatic case. Bottom panel: nonadiabatic case.}
\label{FigWCDM}
\end{figure}

For the dark energy models presented in section \ref{sec:de} we calculate their relative magnitude of the ISW effect in comparison to the $\Lambda$CDM model. Positive (negative) values of $Q$ implies that the model under investigation produces more (less) ISW effect than the standard cosmology. This procedure has been introduced in Ref. \cite{Dent:2008ek}. In Fig. \ref{FigWCDM} we present the results for the $w$CDM model. In this case, the free parameters are $\Omega_{m0}$ and $w=w_{0}$. We calculate the $Q$-values in this free parameters space. For the upper (lower) panel the contours are obtained for the adiabatic (nonadiabatic) dark energy model. In Figs. \ref{FigWCDM}, 2 and 3 the constant $Q$ lines obey the following notation:
\begin{itemize}
\item $Q=0$: Solid line;
\item $\left| Q \right| = 10\%$: Dashed line;
\item $\left| Q \right| = 20\%$: Dashed-dotted line;
\item $\left| Q \right| = 40\%$: Dotted line.
\end{itemize}
The blue contours are the $1\sigma$ and $2\sigma$ contours of confidence level obtained through a joint statistical analysis using the data from SNe Ia (JLA) \cite{Betoule:2014frx}, $H_{0}$ \cite{Riess:2016jrr}, Cosmic Chronometers \cite{Moresco:2012jh,Stern:2009ep,Simon:2004tf} and BAO \cite{Ross:2014qpa,Anderson:2013zyy,Font-Ribera:2013wce}. The statistical analysis was performed using the numerical codes CLASS \cite{Blas:2011rf} and MontePython \cite{Audren:2012wb}. The results of this statistical analysis are shown in Table \ref{TAB}.

\begin{table}
\centering
\begin{tabular}{lcccc} 
\hline\hline 
                  & $\Lambda$CDM                 & $w$CDM                       & CPL                          & WL \\ \hline\hline
$\Omega_{\rm m0}$ & $0.2973_{-0.016}^{+0.015}$   & $0.2982_{-0.016}^{+0.015}$   & $0.2982_{-0.017}^{+0.016}$   & $0.2963_{-0.016}^{+0.015}$ \\ 
$H_{0}$           & $70.23_{-1.4}^{+1.3}$        & $71.14_{-1.6}^{+1.6}$        & $71.17_{-1.6}^{+1.6}$        & $71.53_{-1.4}^{+1.4}$ \\ 
$w_{0}$           & *                            & $-1.003_{-0.050}^{+0.050}$   & $-0.9989_{-0.076}^{+0.077}$  & $-1.034_{-0.053}^{+0.054}$ \\ 
$w_{1}$           & *                            & *                            & $-0.01271_{-0.36}^{+0.37}$   & $0.07671_{-0.110}^{+0.190}$ \\ 
$\chi^{2}_{min}$  & $698.064$                    & $697.774$                    & $697.792$                    & $697.920$ \\ \hline
$\alpha$          & $0.1413_{-0.0067}^{+0.0067}$ & $0.1413_{-0.0067}^{+0.0067}$ & $0.1414_{-0.0068}^{+0.0068}$ & $0.1415_{-0.0068}^{+0.0067}$ \\
$\beta$           & $3.105_{-0.084}^{+0.082}$    & $3.105_{-0.084}^{+0.082}$    & $3.105_{-0.085}^{+0.081}$    & $3.108_{-0.084}^{+0.081}$ \\
$M$               & $-19.04_{-0.046}^{+0.047}$   & $-19.01_{-0.047}^{+0.049}$   & $-19.01_{-0.047}^{+0.048}$   & $-19.00_{-0.041}^{+0.042}$ \\
$\Delta_{M}$      & $-0.070_{-0.024}^{+0.024}$   & $-0.070_{-0.024}^{+0.024}$   & $-0.070_{-0.024}^{+0.024}$  & $-0.070_{-0.024}^{+0.024}$ \\
\hline\hline
\end{tabular}
\caption{Result of the joint statistical analysis using data from SNe Ia (JLA)+$H_{0}$+CC+BAO for all DE parameterizations (for comparison purposes, the $\Lambda$CDM model is also included). The parameters $\alpha$, $\beta$, $M$ and $\Delta_{M}$ are the nuisance parameters from SNe Ia (JLA) data.}
\label{TAB}
\end{table}

It is worth noting that in the top panel of Fig. \ref{FigWCDM}, the case of adiabatic dark energy perturbations, the allowed parameters at $2\sigma$ (light-blue region) can lead to a $\sim 40\%$ difference in the ISW effect. The bottom panel of this figure, the nonadiabatic dark energy perturbations, the $2\sigma$ region is within the $10 \%$ departure of the ISW effect in comparison with the $\Lambda$CDM model.

\begin{figure}
\includegraphics[width=0.45\textwidth]{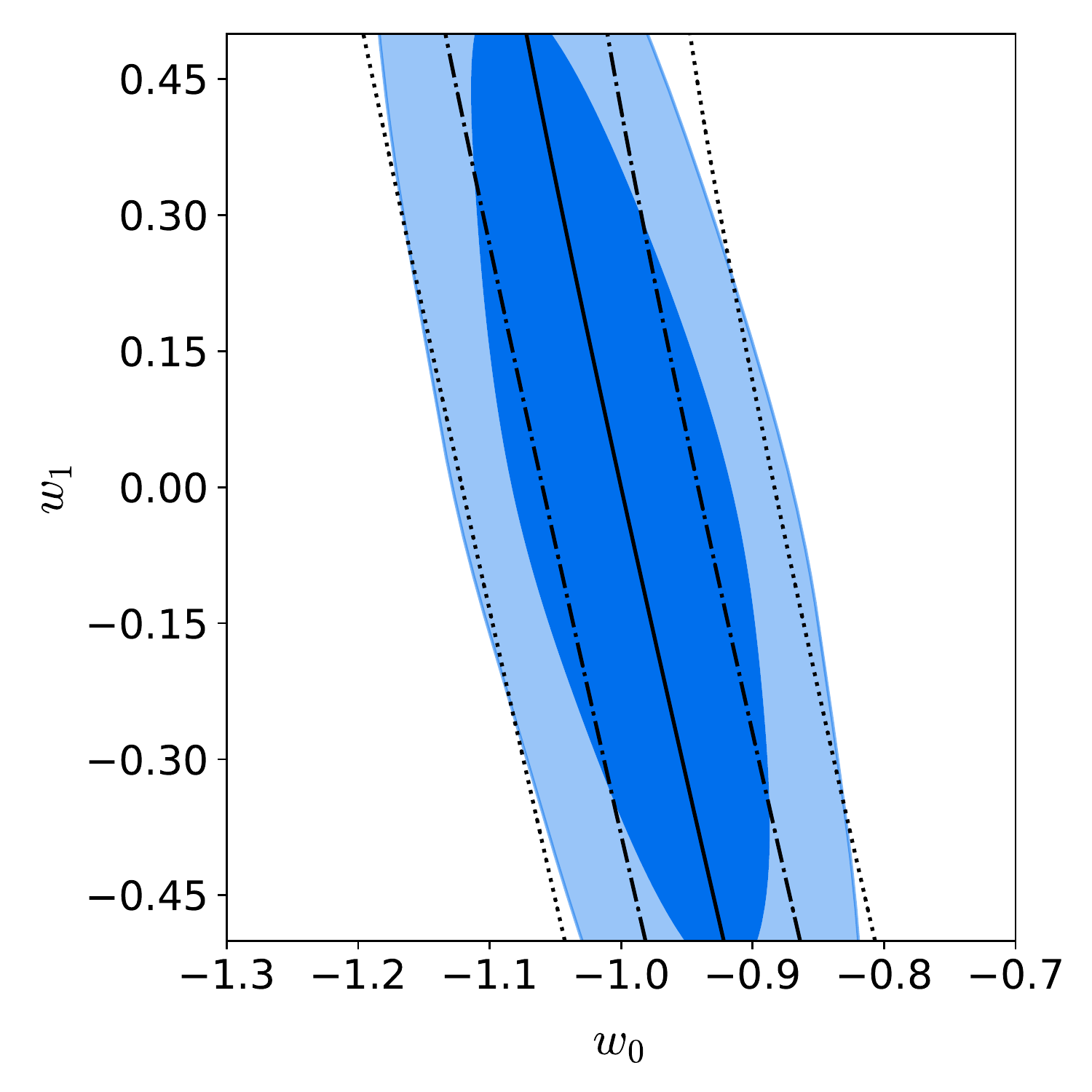}
\includegraphics[width=0.45\textwidth]{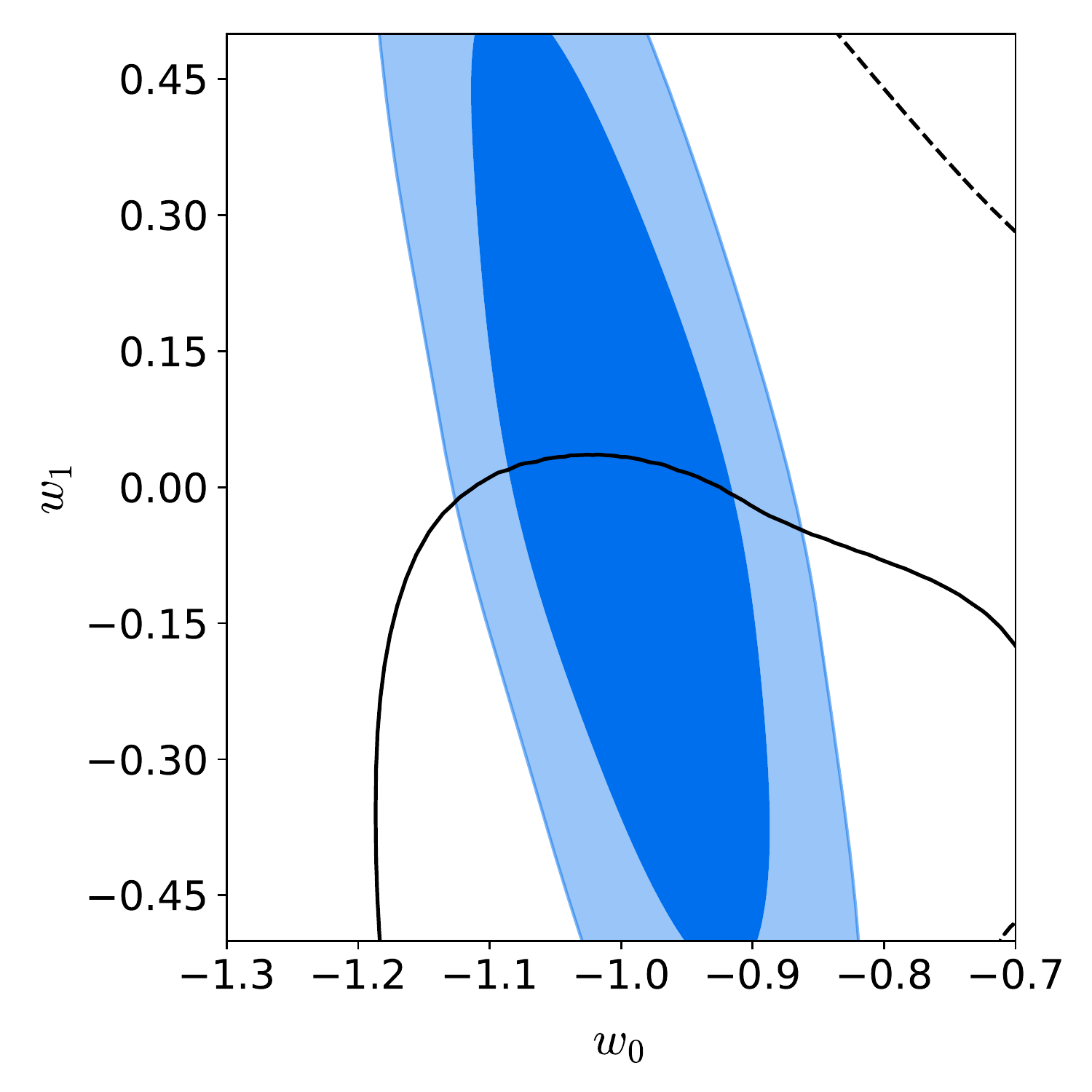}
\caption{Joint Constraints and constant Q-values in terms of the CPL equation of state parameters. Top panel: adiabatic case. Bottom panel: nonadiabatic case.}
\label{FigCPL}
\end{figure}

The result for the CPL parameterization is shown in Fig. \ref{FigCPL}. The same color scheme used in Fig. \ref{FigWCDM} is used. For the Wetterich-logarithmic model results are presented in Fig. 3.

\begin{figure}
\includegraphics[width=0.45\textwidth]{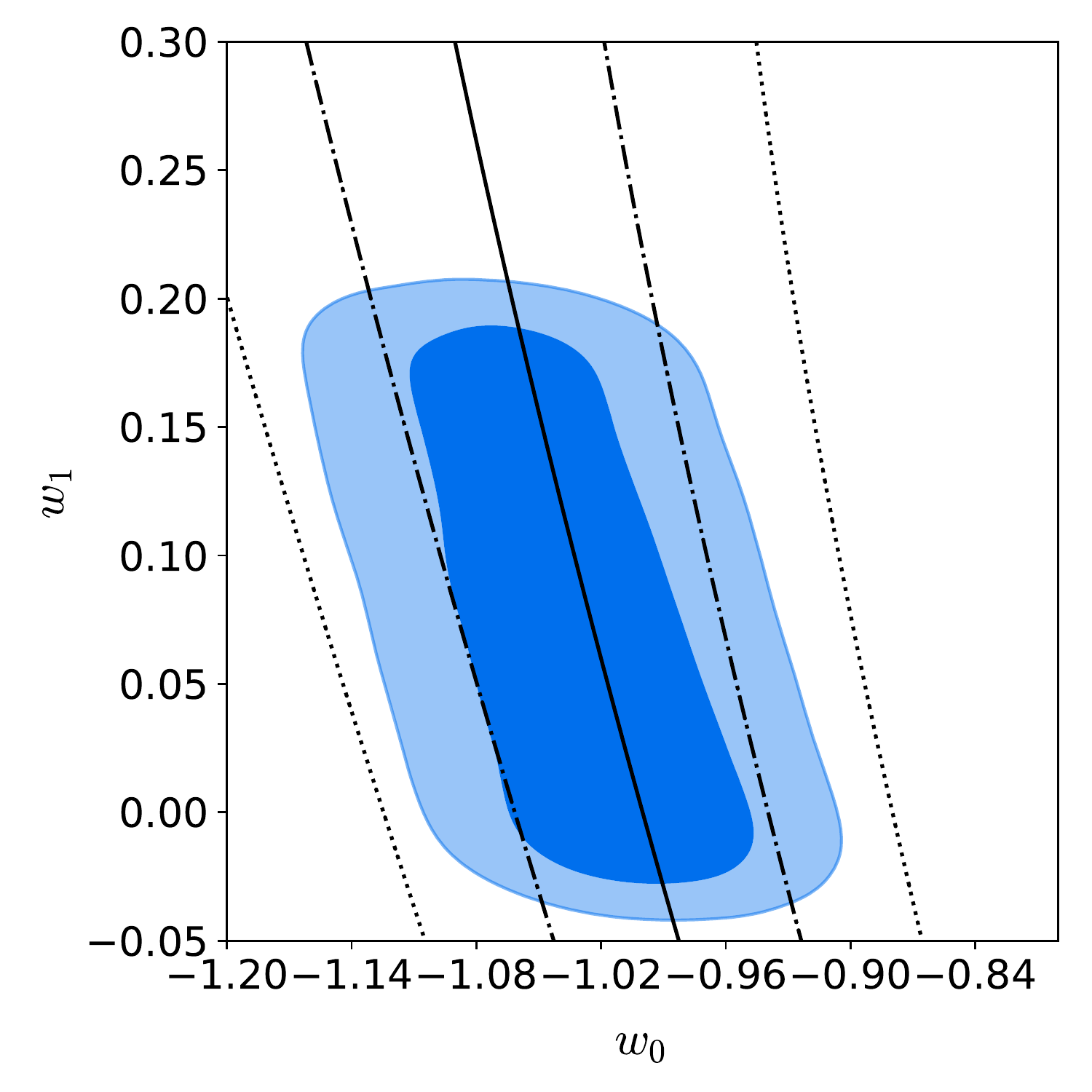}
\includegraphics[width=0.45\textwidth]{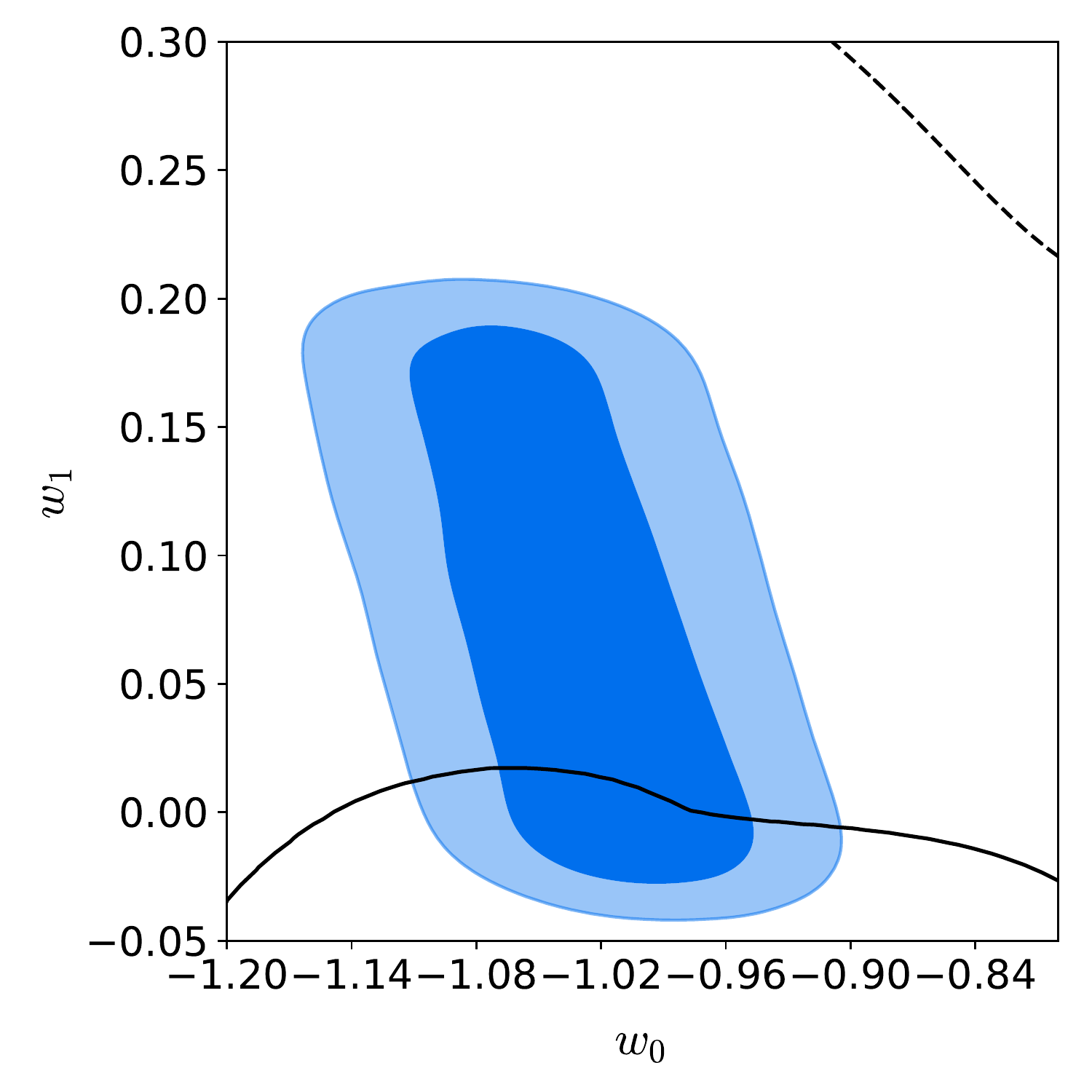}
\caption{Joint Constraints and constant Q-values in terms of the WL equation of state parameters. Top panel: adiabatic case. Bottom panel: nonadiabatic case.}
\label{FigWL}
\end{figure}

Figs. \ref{FigWCDM}, \ref{FigCPL} and \ref{FigWL} share similar results, from which we can conclude that the nonadiabatic dark energy models have their predictions concerning the magnitude of the ISW effect much closer to the $\Lambda$CDM model. This result confirms the findings of Ref. \cite{Velten:2017mtr}. 

\section{\label{sec:Results} Confronting scalar perturbations with CMB-LSS cross correlation data (the ISW effect)}\label{crossISWLSS}

In this section we promote a more robust analysis about the impact of nonadiabatic dark energy cosmologies on the late time ISW effect. We calculate the cross-correlation of CMB with LSS data. In order to perform such analysis it is worth noting that we have to assess the matter growth $\delta_m$ rather than the total single fluid density contrast. In order to correctly calculate the matter growth we obtain an equation for the evolution of matter density perturbation $\delta_m$ as a function of $\Phi$. By adapting Eqs. (\ref{cont}) and (\ref{Euler}) for pressureless matter i.e., $w\rightarrow w_m=0$ (and making $\Delta \rightarrow \delta_{\rm m}$) we find
\begin{equation}
a^2 \delta^{\prime\prime}_m+\left(\frac{a H^{\prime}}{H}+3\right)a\delta^{\prime}_m+\frac{k^2 H^2_0}{a^2 H^2(a)}\Phi=0.
\label{deltam}\end{equation}
Therefore, as done in the last section, we can numerically solve Eqs. (\ref{Pot1}), (\ref{Pot2}) and (\ref{Pot3}) for the potential $\Phi$ and use this solution as the source to calculate the evolution of $\delta_m$ with Eq. (\ref{deltam}).

In order to compute the cross-correlation between CMB and LSS we also need to describe the evolution of the observed galaxy contrast $\delta_g$ on the line of sight. This quantity depends on the survey design and is obtained from
\begin{equation}\label{deltagal}
\delta_g = \int^{\chi_d}_0 b(z) \frac{dN}{dz} H(a) \delta_{m}(a) d\chi,
\end{equation}
where in the above equation we have used as the dynamical variable the comoving distance
\begin{equation}
\chi(a)=\int^1_a \frac{da}{a^2 H(a)}.
\end{equation}
The integration in Eq. (\ref{deltagal}) runs until the comoving distance of the decoupling epoch $\chi_d$. Also, $\delta_{m}$ is the total matter linear contrast, and $b(z)$ is the bias function between the galaxy overdensity $\delta_g$ and $\delta_m$. Galaxy surveys provide the distribution of observed galaxies along the line of sight ($dN/dz$) in a model-independent way. In this work we use data from the NRAO VLA Sky Survey (NVSS) \footnote{http://www.cv.nrao.edu/nvss/} and the Wide-field Infrared Survey (WISE) \footnote{http://wise.ssl.berkeley.edu/} catalogues. The NVSS covers the entire north sky of $-40$ deg declination in one band. The resulting catalogue of discrete sources has about one million objects. More details can be found in Ref.\cite{Condon:1998iy}. The WISE survey scans the entire sky in four frequency bands. It reachs up to $z \sim 1$, but with an average redshift $z=0.3$. The data used in this work has been taken from the analysis recently performed in Ref. \cite{Ferraro:2014msa}.

The galaxy distribution of the NVSS catalogue is given by the function 
\begin{equation}
\left[b(z)\frac{dN}{dz}\right]_{NVSS} =b_{eff}\frac{\alpha^{\alpha+1}}{z_{\star}^{\alpha+1}\Gamma(\alpha)}z^{\alpha} {\rm exp} \left[-\frac{\alpha z}{z_{\star}}\right],
\end{equation}
where the it has been fixed the parameters $b_{eff}=1.98$, $z_{\star}=0.79$ and $\alpha=1.18$\cite{Ho:2008bz}.

The WISE galaxy distribution $\left[dN / dz\right]_{WISE}$ has been obtained numerically \cite{Yan:2012yk}. Following Ref. \citep{Ferraro:2014msa} we adopt a constant bias $\left[b(z)\right]_{WISE}=1.41$. We will limit our analysis to constant bias models since the use of time-dependent $b(z)$ bias does not change our main conclusion.

Now, in order to compute the ISW effect signal from the matter growth we make use of the Poisson equation (in the quasi-static approximation) in the first equality of Eq. {\ref{DeltaTISW}}. Thus, with such changes described above, the ISW signal reads
\begin{eqnarray}\label{deltaTT}
\frac{\Delta T}{T}^{ISW}=\frac{\delta_{m}(a=1)}{k^2}\int^{\chi_d}_0 a^2 H(a) \frac{d\chi_{m} (a)}{da} d\chi,
\end{eqnarray}
where we new function $\chi_m(a)$ is defined according to
\begin{eqnarray}\label{defQ}
\chi_{m}(a) &=& a^2 \rho_{m}(a) \frac{\delta_m(a)}{\delta_{m}(a=1)}.
\end{eqnarray}

By cross correlating the galaxy distribution overdensity with the ISW signal we combine (\ref{deltaTT}) and (\ref{deltagal}). Thus, the multipole coefficients for the cross-spectrum are calculated with the expression
\begin{equation}
C^{Tg}_l =  \int^{a_d}_{1} \frac{W_T (a)\, W_g (a)}{a^2 H(a)} \frac{P(k=l/\chi)}{l^2} da.
\label{Ctg}\end{equation}
In the above result the weight functions have been defined as
\begin{eqnarray}
W_T\,(a) &=& a^2 H(a) \frac{d\chi_m(a)}{da}, \\
W_g\,(a) &=& H(a) b(z)\frac{dN}{dz} \frac{\delta_m(a)}{\delta_{m}(a=1)}.
\end{eqnarray}
In expression (\ref{Ctg}) the power spectrum $P(k)=\left|\delta_{m}(k)\right|^2$ is introduced. We can calculate it via the standard result $P(k) = P_0 k^{n_s} T^2(k)$. The shape of the power spectrum is set by the transfer function $T(k)$. We contribution to $P(k)$ from the primordial inflationary spectrum is encoded in the spectral index $n_{s}$. We adopt a Harrison-Zeldovich approximation $n_{s}=1$. For the transfer function we use the BBKS fit \cite{Bardeen:1985tr}, which is accurate enough to the data,
\begin{eqnarray}
&&\left[T(x=k/k_{eq})\right]_{BBKS}=\frac{ln[1+0.171x]}{(0.171x)}\times \\ \nonumber
&&\left[1 + 0.284x + (1.18x)^2 + (0.399x)^3 + (0.490x)^4\right]^{-0.25}.
\end{eqnarray}
In the above formula $h = H_0/(100$ km/s-Mpc) and we have set the present radiation density as $\Omega_{R0} = 4.15 \times 10^{-5} h^{-2}$ in order to compute the equality between matter and radiation. The procedure described above has been performed in details in Ref. \cite{Velten:2015qua}.

\begin{figure}[!t]
\includegraphics[width=0.5\textwidth]{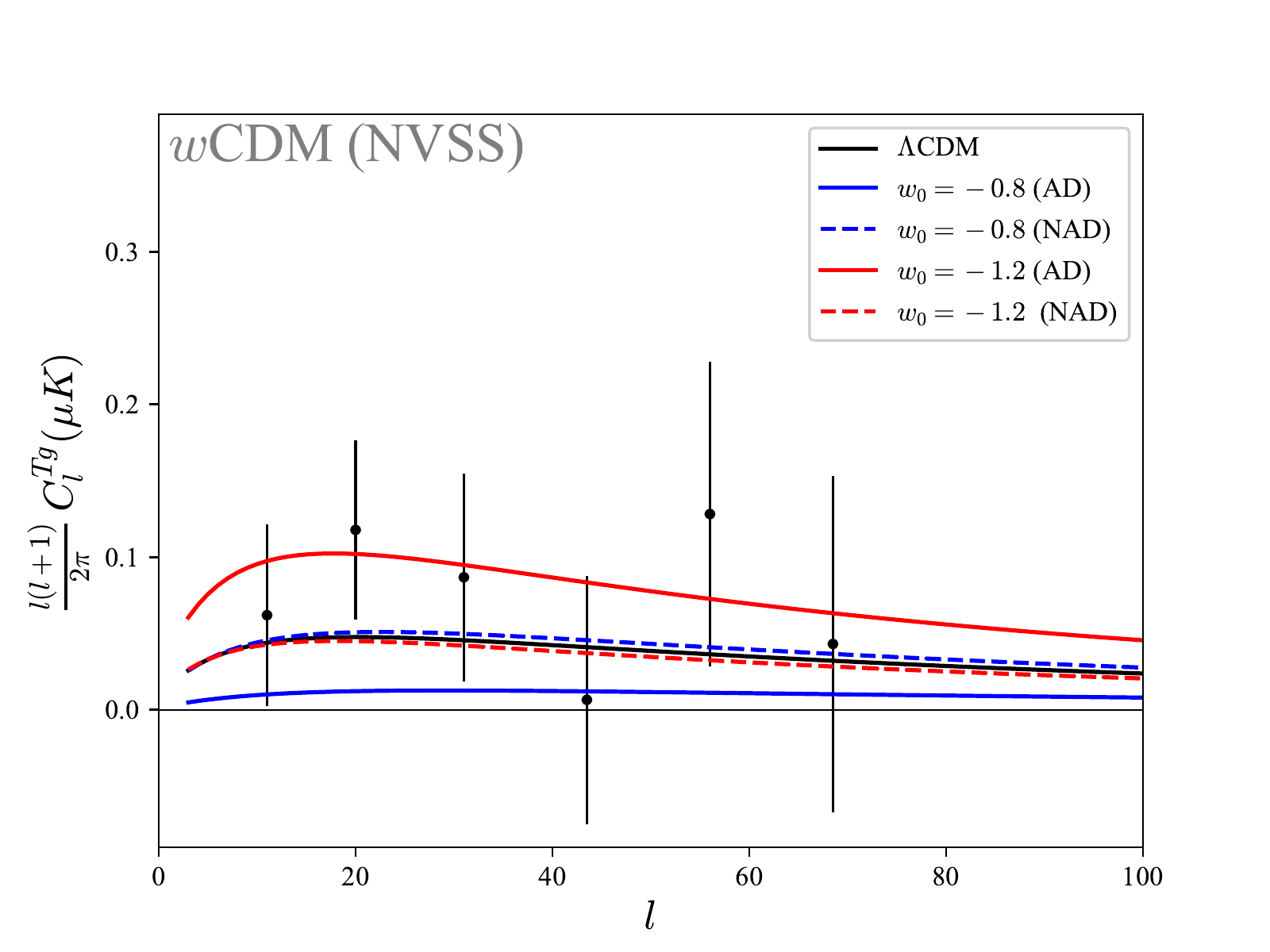}
\includegraphics[width=0.5\textwidth]{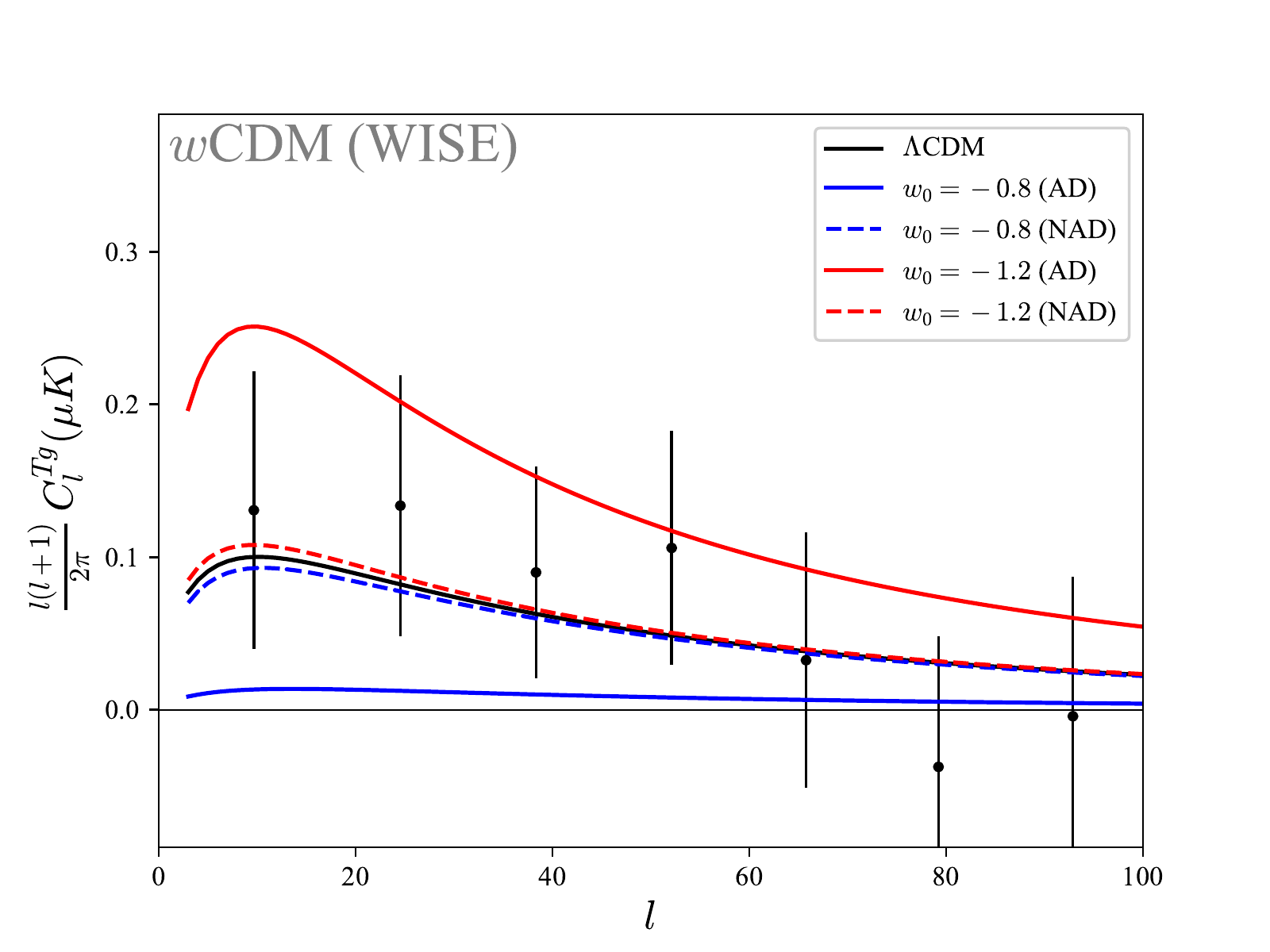}
\caption{The $C^{Tg}_{l}$ power spectrum for the $w$CDM. Top panel: NVSS data. Bottom panel: WISE data.}
\label{FigWCDMCtg}
\end{figure}

\begin{figure}[!t]
\includegraphics[width=0.5\textwidth]{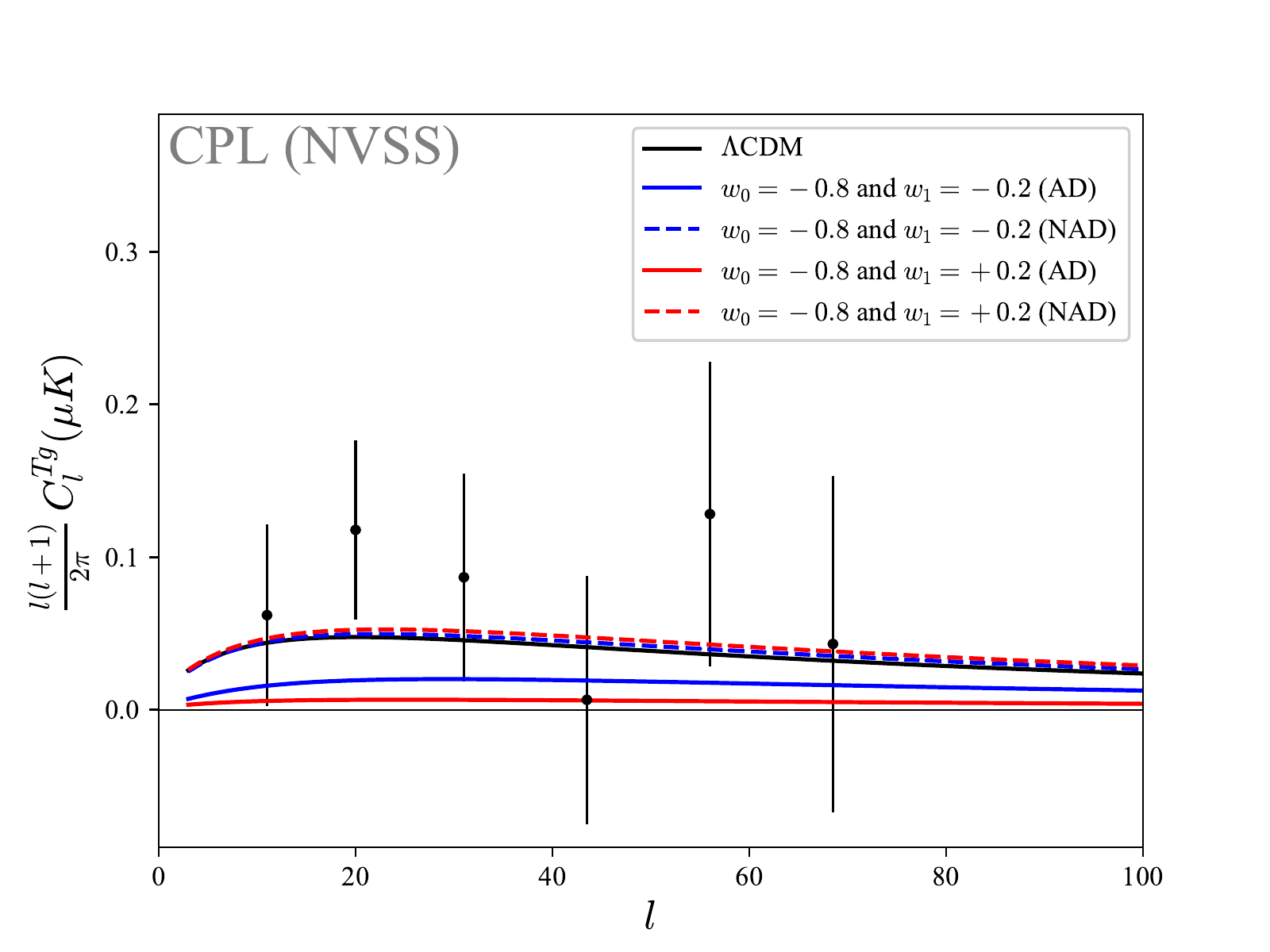}
\includegraphics[width=0.5\textwidth]{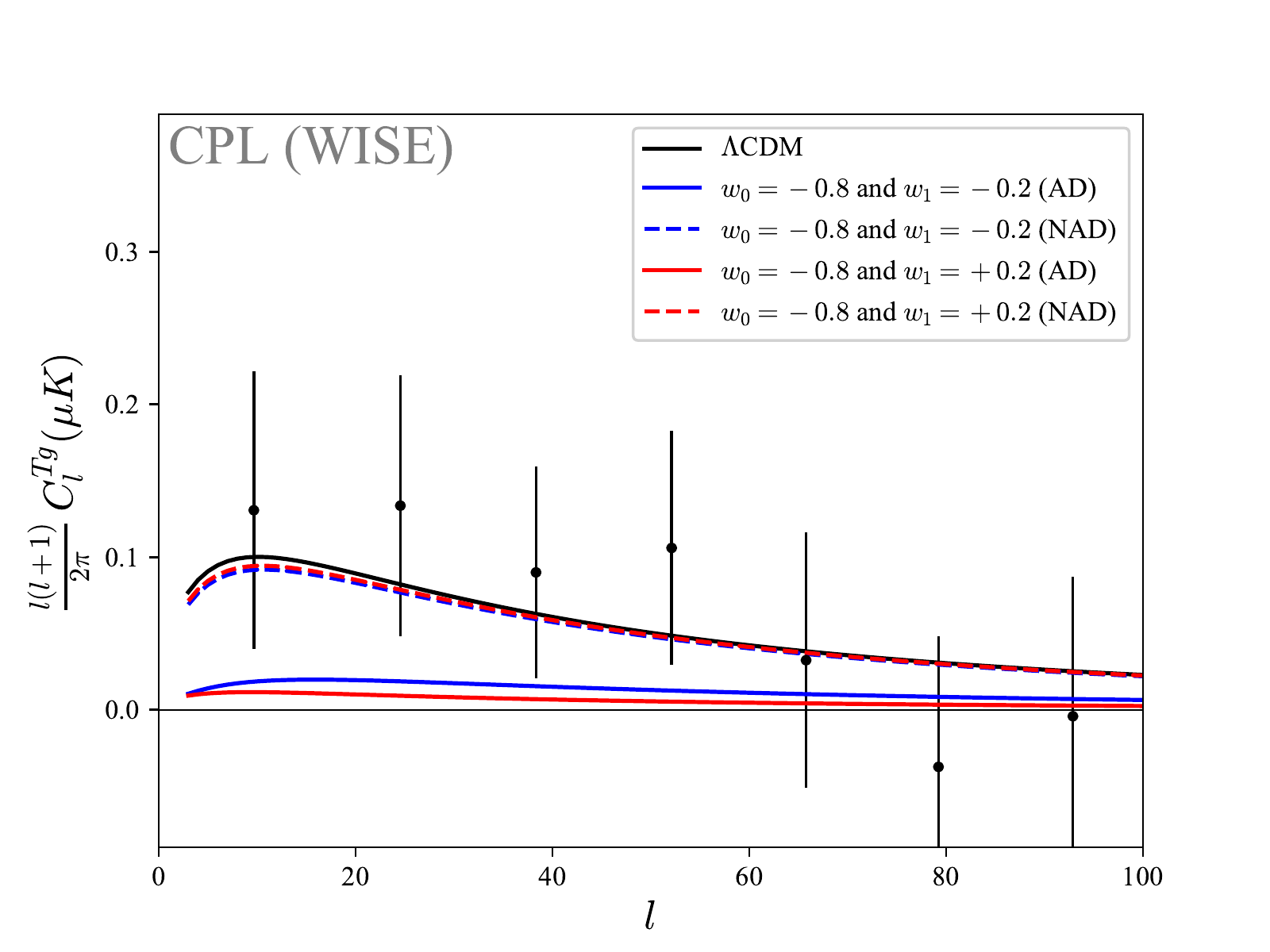}
\caption{The $C^{Tg}_{l}$ power spectrum for the CPL model with $w_{0}=-0.8$. Top panel: NVSS data. Bottom panel: WISE data.}
\label{FigCPLCtg1}
\end{figure}

\begin{figure}[!t]
\includegraphics[width=0.5\textwidth]{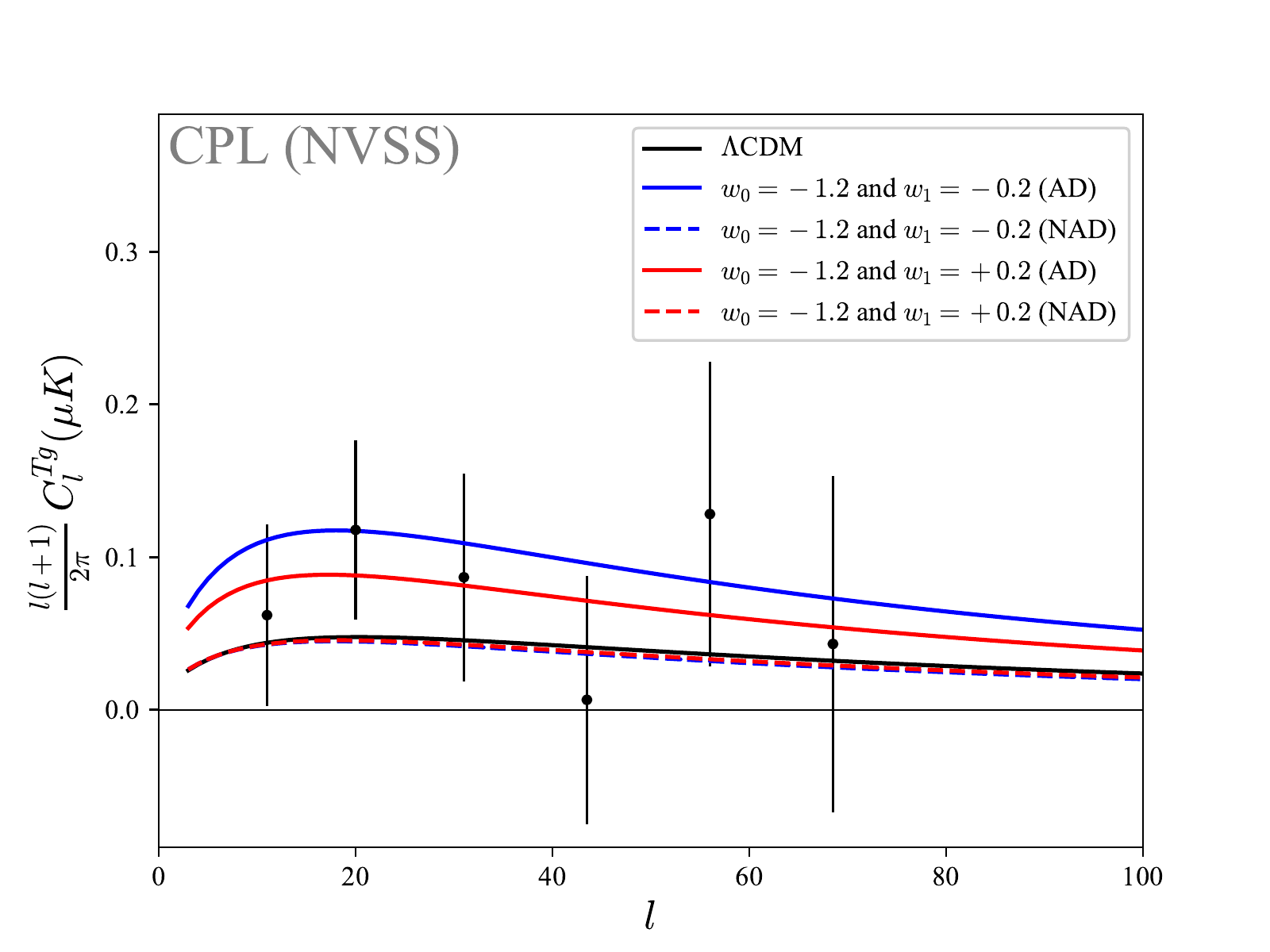}
\includegraphics[width=0.5\textwidth]{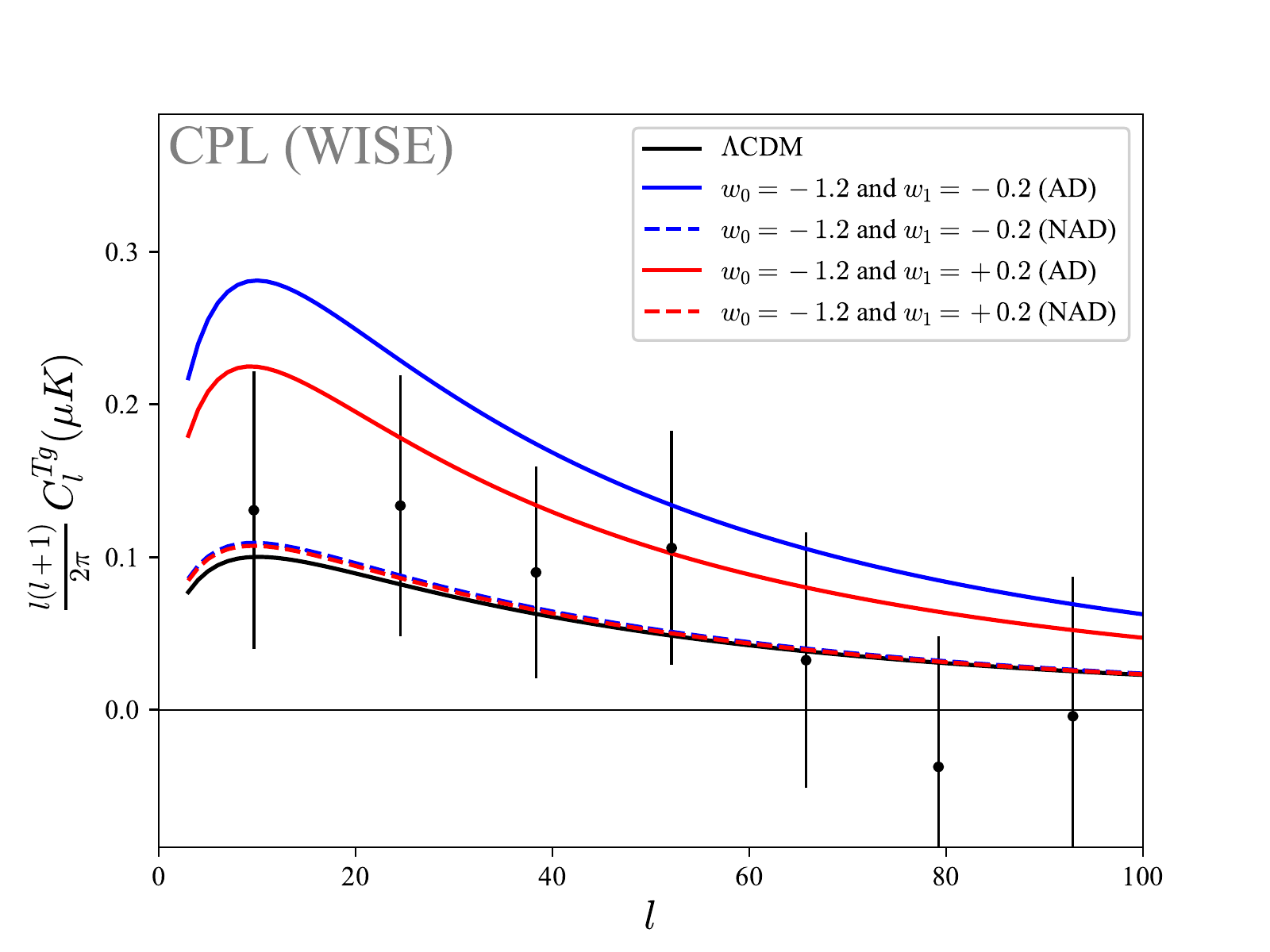}
\caption{The $C^{Tg}_{l}$ power spectrum for the CPL model with $w_{0}=-1.2$. Top panel: NVSS data. Bottom panel: WISE data.}
\label{FigCPLCtg2}
\end{figure}

\begin{figure}[!t]
\includegraphics[width=0.5\textwidth]{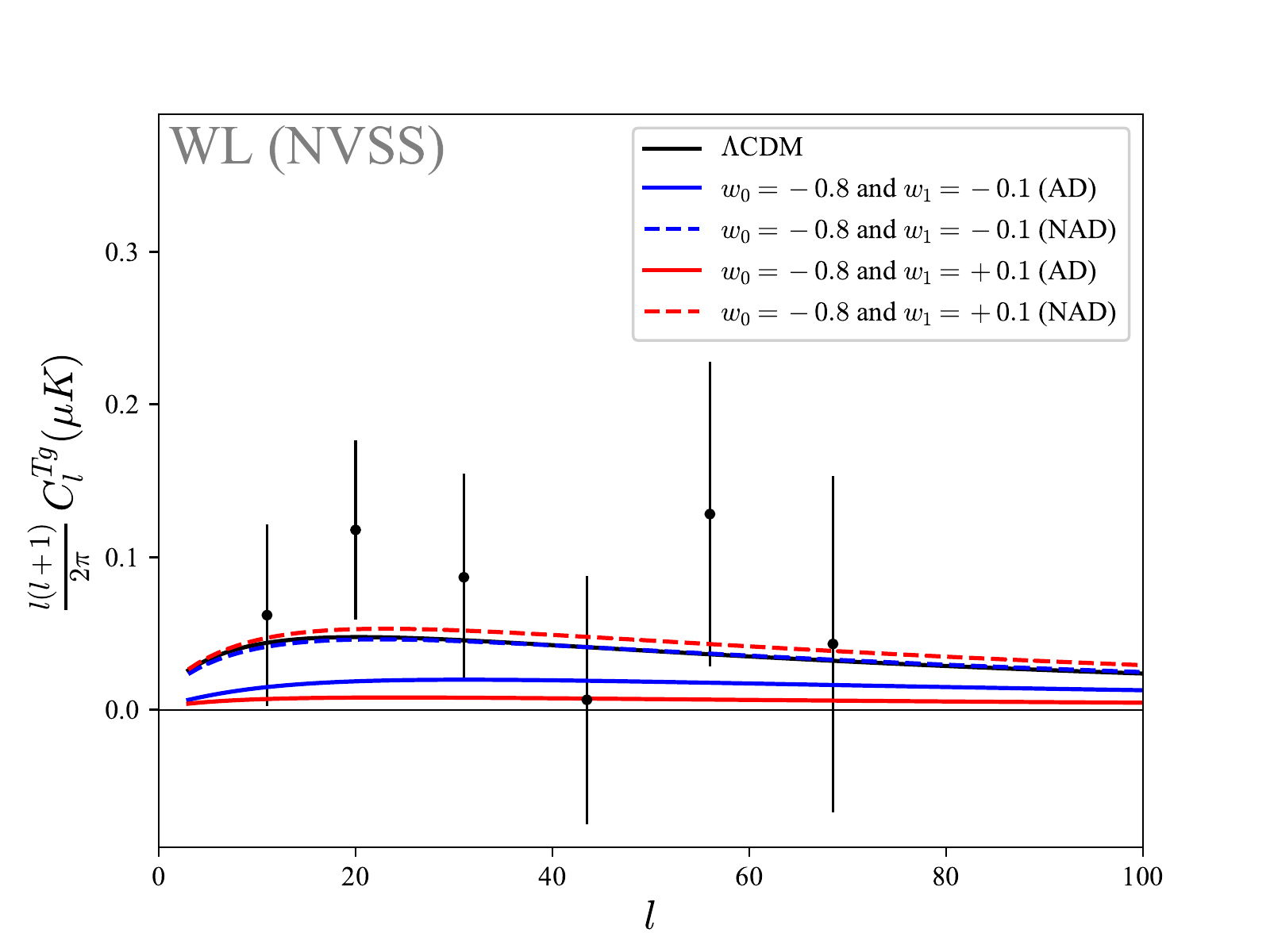}
\includegraphics[width=0.5\textwidth]{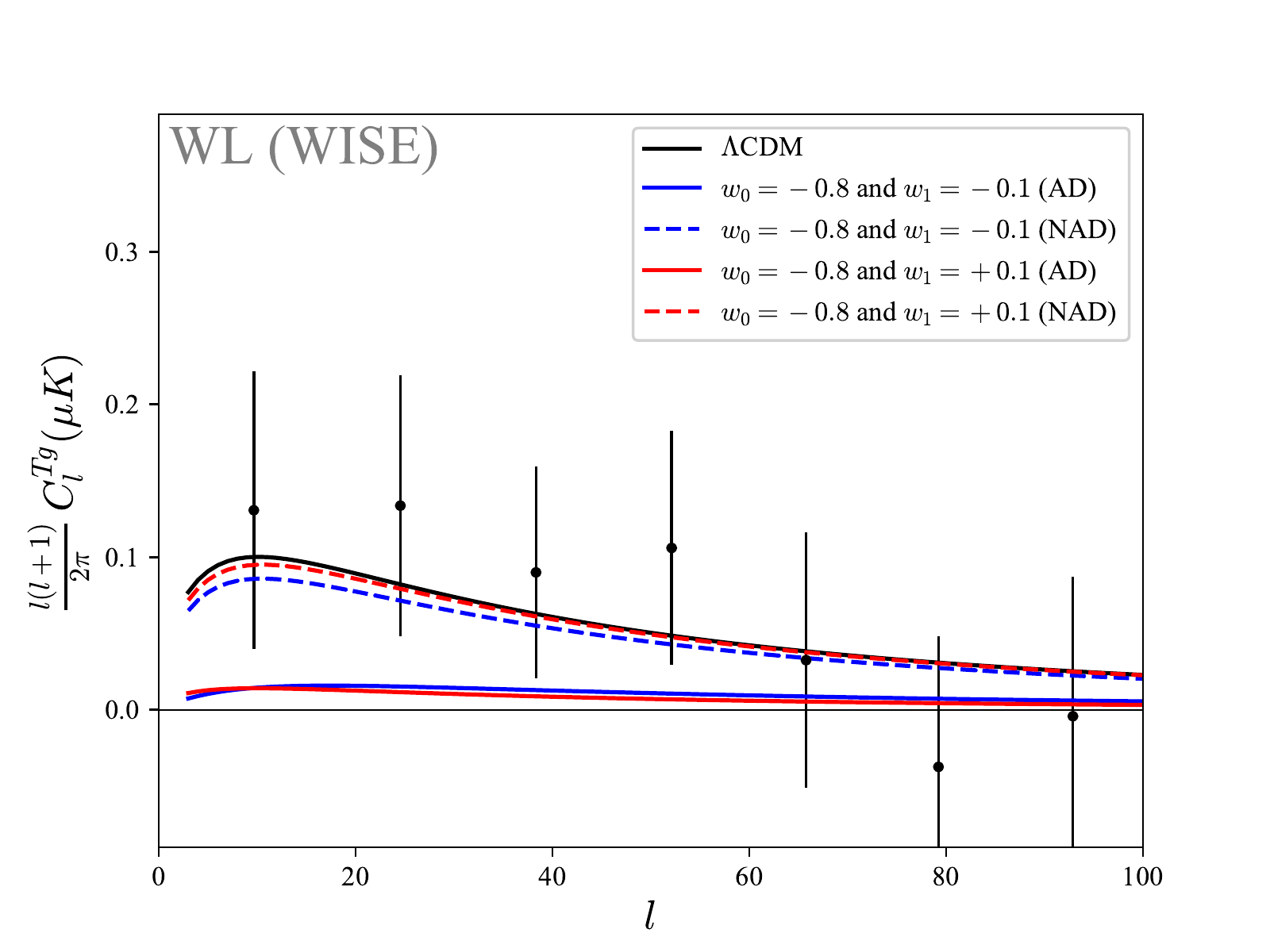}
\caption{The $C^{Tg}_{l}$ power spectrum for the WL model with $w_{0}=-0.8$. Top panel: NVSS data. Bottom panel: WISE data.}
\label{FigWLCtg1}
\end{figure}

\begin{figure}[!t]
\includegraphics[width=0.5\textwidth]{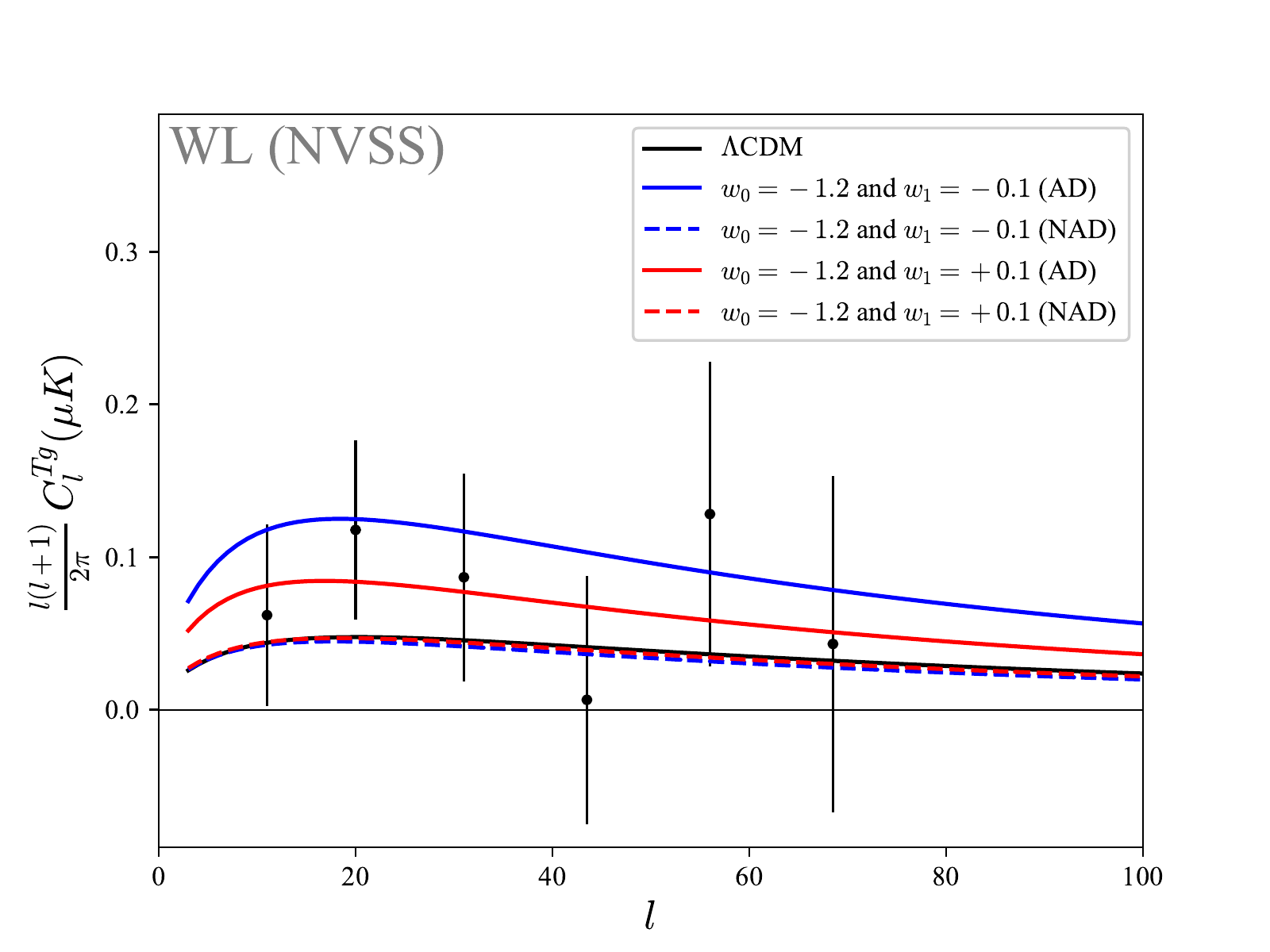}
\includegraphics[width=0.5\textwidth]{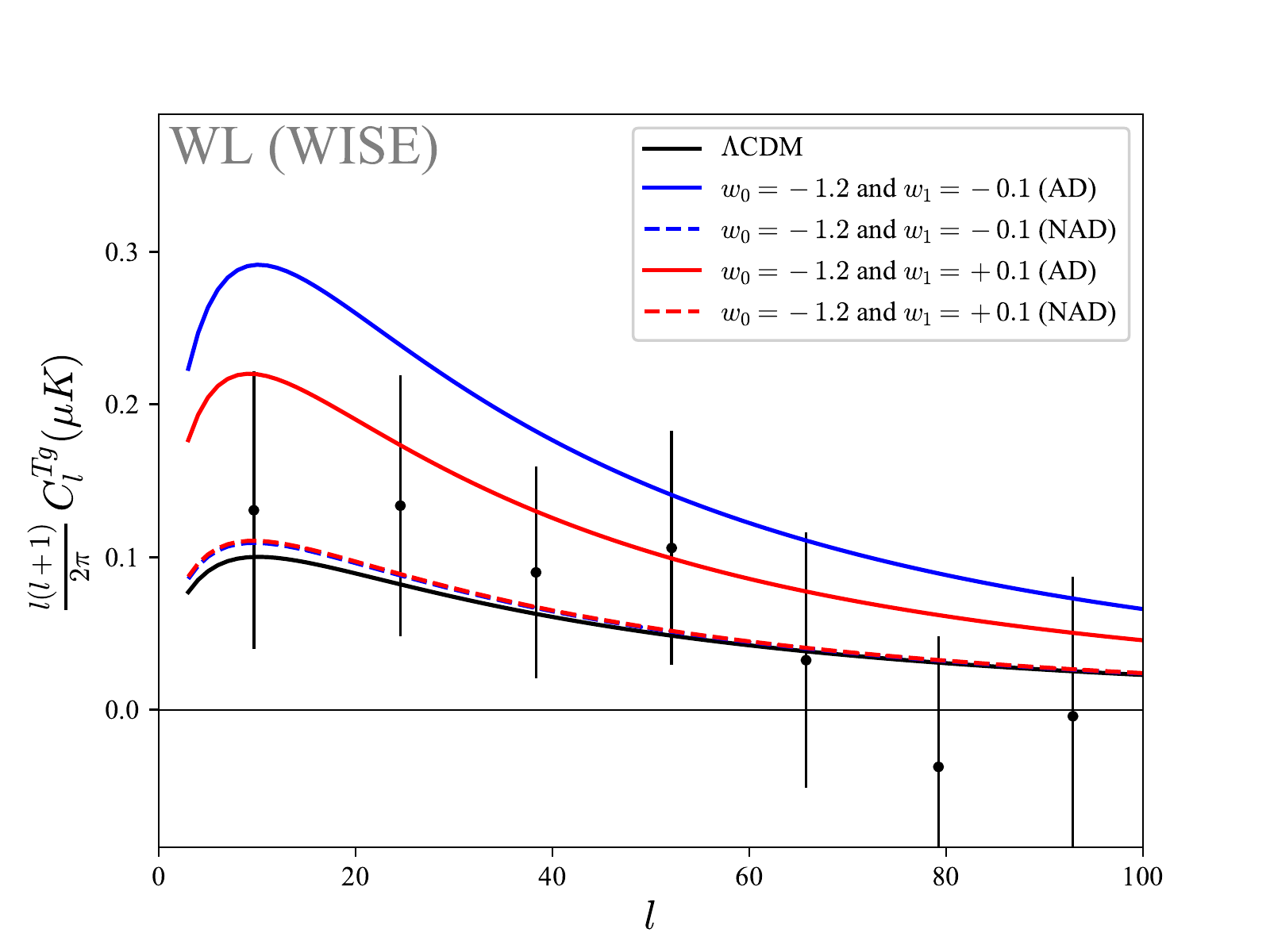}
\caption{The $C^{Tg}_{l}$ power spectrum for the WL model with $w_{0}=-0.8$. Top panel: NVSS data. Bottom panel: WISE data.}
\label{FigWLCtg2}
\end{figure}

The resulting cross-correlation spectrum for the dark energy models studied here are shown in Figs. \ref{FigWCDMCtg} - \ref{FigWLCtg2}. In both figures the top panels correspond to the use of the NVSS data while the bottom panels to the WISE data. In such figures the solid black represents the standard $\Lambda$CDM model. Adiabatic (AD) models are plotted in solid lines. Nonadiabatic models (NAD) in dashed lines. The equation of state parameter values adopted in each curve of Figs. \ref{FigWCDMCtg} to \ref{FigWLCtg2} are demonstrated in the inset in the upper right region of each panel. 

In Fig. \ref{FigWCDMCtg} we study the $w$CDM model. Quintessence dark energy with $w_0=-0.8$ is plotted in the blue lines and the phantomic case $w_0=-1.2$ in the red lines. Indeed, as seen in Table I these values are not preferred by current available data. The cross-spectrum data is not able to constrain $w_0$ with such precision but a qualitative inspection of the curves in Fig. \ref{FigWCDMCtg} shows that the solid blue and solid red lines clearly departures from the $\Lambda$CDM model. The main aspect which we want to stress out in this work is how close the nonadiabatic models (dashed) are to the standard cosmology.

The results for the CPL model are shown in Figs. \ref{FigCPLCtg1} and \ref{FigCPLCtg2}. The WL model in Figs. \ref{FigWLCtg1} and \ref{FigWLCtg2}. It is worth noting that the same pattern concerning the quasi-degeneracy between nonadiabatic dark energy models and the $\Lambda$CDM model persists in these cases.

Although the $C^{Tg}_{l}$ data is not accurate enough mainly due to the cosmic variance at low multipoles the $\Lambda$CDM as calculated here provides a consistent (statistically) fit to the data. There are however claims in the literature about a possible inconsistent matching between the observable ISW effect and the predictions of the $\Lambda$CDM model
\cite{Cai:2016rdk}. Also, recent investigation points out to a tension between the expected ISW of a $\Lambda$CDM universe and the signal obtained from cosmological simulations \cite{Beck:2018owr}.

\section{\label{sec:Conclusion}Conclusions}

In this work we have analyzed dark energy models by taking into account the possible nonadiabatic effects such that the intrinsic and relative entropy perturbations. Obviously such effects do not manifest at the background level but are relevant for the perturbations. We have focused on scalar perturbations and the impact of the nonadiabaticity of dark energy models on the gravitational potential evolution and the CMB-galaxy cross-spectrum. Both analysis are related to the integrated Sachs-Wolf effect. Our findings can be interpreted in the same way as in Ref. \cite{Velten:2017mtr}, i.e., while adiabatic dark energy models differs substantially from the standard $\Lambda$CDM cosmology if the equation of state parameter departures from the constant value $w_{DE}=p_{DE}/\rho_{DE}=-1$, there seems to exist a sort of quasi-degeneracy between nonadiabatic dark energy models and the $\Lambda$CDM model. 

Since this apparent degeneracy is completely characterized at first order in liner perturbations the future investigations should focus on whether it persists at second order or within the nonlinear regime of perturbations.

\section*{Acknowledgments}
We acknowledge fruitful discussions with Winfried Zimdahl. HV thanks CNPq (Brazil) and FAPES (Brazil) for partial support. The work of REF is supported by CNPq. The work of RvM is supported by CNPq (PDJ program). The work of SG is supported by (UFES/CNPq).

\end{document}